\documentclass[3p,times,procedia]{elsarticle}
\usepackage{nupha_ecrc}

\volume{00}

\firstpage{1}

\journalname{Nuclear Physics A}

\runauth{}

\jid{nupha}

\jnltitlelogo{Nuclear Physics A}

\usepackage{amssymb}

 \usepackage{lineno}

\usepackage[figuresright]{rotating}
\biboptions{sort&compress}

\begin{document}

\begin{frontmatter}



\dochead{XXVIIIth International Conference on Ultrarelativistic Nucleus-Nucleus Collisions\\ (Quark Matter 2019)}

\title{Measurement of cumulants of conserved charge multiplicity distributions in Au+Au collisions from the STAR experiment}


\author{Ashish Pandav for the STAR Collaboration}
\address{School of Physical Sciences, National Institute of Science Education and Research, HBNI, Jatni-752050, INDIA}

\begin{abstract}
 We report the collision-centrality dependence of cumulants  of event-by-event net-proton, net-charge and net-kaon distributions in Au+Au collisions for center-of-mass energy $\sqrt{s_{NN}}$ = 54.4 GeV from the STAR experiment. Strong collision-centrality dependence is observed for cumulants ($C_{n}, n$ $\leq$ 4) of the net-particle distributions. The cumulant ratios $C_{3}/C_{2}$, $C_{4}/C_{2}$ and $C_{6}/C_{2}$ exhibit a weak collision-centrality dependence. The $C_{6}/C_{2}$ of net-proton and net-charge distributions for most central gold nuclei collisions at $\sqrt{s_{NN}}$ = 54.4 GeV show positive sign while they remain negative  at 200 GeV for the same collision system. To understand effects of acceptance and baryon number conservation, the measurements are compared to expectations from the UrQMD and HIJING models calculated within the STAR detector acceptance.
\end{abstract}


\end{frontmatter}


\section{Introduction}
\label{}
One of the major goals of heavy-ion collision experiments is to explore the Quantum Chromodynamics (QCD) phase diagram and search for the QCD critical point. Lattice QCD calculations have shown that for vanishing baryonic chemical potential ($\mu_{B}$), the nature of phase transition between quark-gluon plasma (QGP) and hadronic matter is a smooth crossover~\cite{Aoki:2006we} whereas QCD-based models suggest this phase transition to be of first order for finite  $\mu_{B}$~\cite{Ejiri:2008xt}. The cumulants of conserved quantities in strong interactions are proposed to be sensitive observables for the search of the QCD critical point and the phase transition between quark-gluon plasma (QGP) and hadronic matter~\cite{Stephanov:2008qz}. The cumulants and their ratios are related to the correlation length of the hot and dense medium formed in the heavy ion collisions and the thermodynamic susceptibilities that are calculable via various QCD-based models and lattice QCD~\cite{Gavai:2010zn,Gupta:2011wh}.

 Cumulants quantify the traits of a distribution, for example, the first- and second-order cumulant ($C_{1}$ and $C_{2}$) are the mean and variance of a distribution whereas the third- and fourth-order cumulants ($C_{3}$ and $C_{4}$) reflect the skewness and kurtosis of a distribution, respectively. Cumulants and their ratios for event-by-event distributions of net-charge, net-kaon and net-proton in collision of gold nuclei were measured by the STAR detector in the phase I of Beam Energy Scan (BES) program at the Relativistic Heavy Ion Collider (RHIC)~\cite{starNC,starNK,starNP_old,starNP,shor_nature}. Non-monotonic dependence on beam energy is observed for the cumulant ratios $C_{3}/C_{2}$ and $C_{4}/C_{2}$ of net-proton distribution in the most central (0-5\%) collisions, which may hint at existence of a possible critical point. The sixth-order cumulants ($C_6$) could also provide insights into the nature of phase transition. Negative $C_6/C_2$ of net-baryon and net-charge distributions are predicted from a QCD-based model for crossover phase transitions, if the chemical freeze-out is close to the chiral phase transition~\cite{negc6:friman}.
 
 \section{Analysis techniques}
 About $\sim$ 550 million events are analysed for obtaining cumulants of net-particle distributions in Au+Au collisions at $\sqrt{s_{NN}}$ = 54.4 GeV. Time Projection Chamber (TPC) and Time-of-Flight (TOF) detectors~\cite{Shao:2005iu} are used to select protons (antiprotons) within $p_{T}$ range 0.4 -- 2.0 GeV/c, and charged pions and kaons within $p_{T}$ range 0.2 -- 1.6 GeV/c. The rapidity coverage $|y|<$ 0.5 is used to select charged particles for measuring net-proton and net-kaon cumulants while psuedorapidity coverage $|\eta|<$ 0.5 is considered for charged particle selection for net-charge fluctuation measurements. The collision centrality is determined from the charged particle multiplicity excluding the particles of interest to avoid the autocorrelation effect~\cite{arghya_auto}. In order to suppress the volume fluctuation effects, centrality bin width correction is applied to the measurement of the cumulants~\cite{bin_width}. Cumulants are correction for finite efficiency and acceptance effects of the detector with the assumption that the distribution of the detector response is binomial~\cite{efcor_luo,Nonaka:2017kko}. For estimation of statistical uncertainties of cumulants and their ratios, delta theorem method and a resampling method called the bootstrap are used \cite{Luo:2011tp,bootstrap}. Systematic uncertainties of the $C_{n}$'s are estimated varying tracking efficiency, track selection and particle identification criteria. 
 
 \section{Results}
 \subsection{Cumulants of net-particle distributions}
 Cumulants up to the $4^{th}$ order of the event-by-event net-proton, net-charge and net-kaon distributions for Au+Au collisions at $\sqrt{s_{NN}}$ = 54.4 GeV as a function of collision centrality (given by the average number of participant nucleons, $<$Npart$>$) are presented in Fig. 1. The statistical and systematic uncertainties on the measurements are shown by red bars and black brackets, respectively. Cumulants of the net-partcle distributions increase from peripheral to central collisions. Cumulants of the net-charge distribution have the largest statistical uncertainties for a given centrality which can be attributed to the larger width of net-charge distributions among all net-particle distributions for a given centrality.
 \begin{figure}[h]
	\hspace*{+0.3cm}
	\includegraphics[scale=0.57]{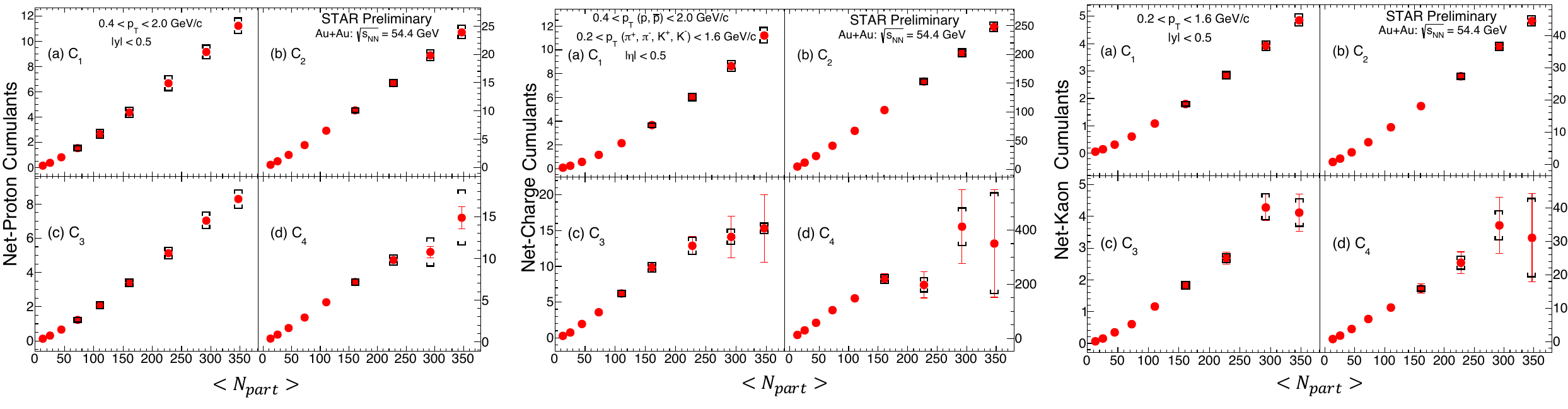}
	\caption{Cumulants ($C_{n}, n$ $\leq$ 4) of net-proton, net-charge, net-kaon distributions as a function of average number of participant nucleons for Au+Au collisions at $\sqrt{s_{NN}}$ = 54.4 GeV.}
	\label{cumu}
\end{figure}
\subsection{Cumulant ratios of net-particle distributions}
Figure 2 shows collision-centrality dependence of cumulant ratios $C_{2}/C_{1}$, $C_{3}/C_{2}$ and $C_{4}/C_{2}$ ($\sigma^{2}/M$, $S\sigma$ and $\kappa\sigma^{2}$ respectively) of net-particle distributions constructed from the measured cumulant values. While $C_{2}/C_{1}$ decreases with collision centrality, the cumulant ratios $C_{3}/C_{2}$ and $C_{4}/C_{2}$ exhibit a weak dependence on collision centrality for net-proton, net-charge and net-kaon distributions. The expectations from the UrQMD and HIJING models are also compared to the measurements~\cite{urqmd,hijing}. The model expectations are inconsistent with the measurements and only qualitatively reproduce the measured centrality dependence of the cumulant ratios. The Skellam baseline for $C_{4}/C_{2}$, which is the expected value of $C_{4}/C_{2}$ under the assumption that protons and antiprotons follow Poisson distribution independently, fails to describe the measured values.
 \begin{figure}[h]
 	\hspace*{+2.6cm}
	\includegraphics[scale=0.51]{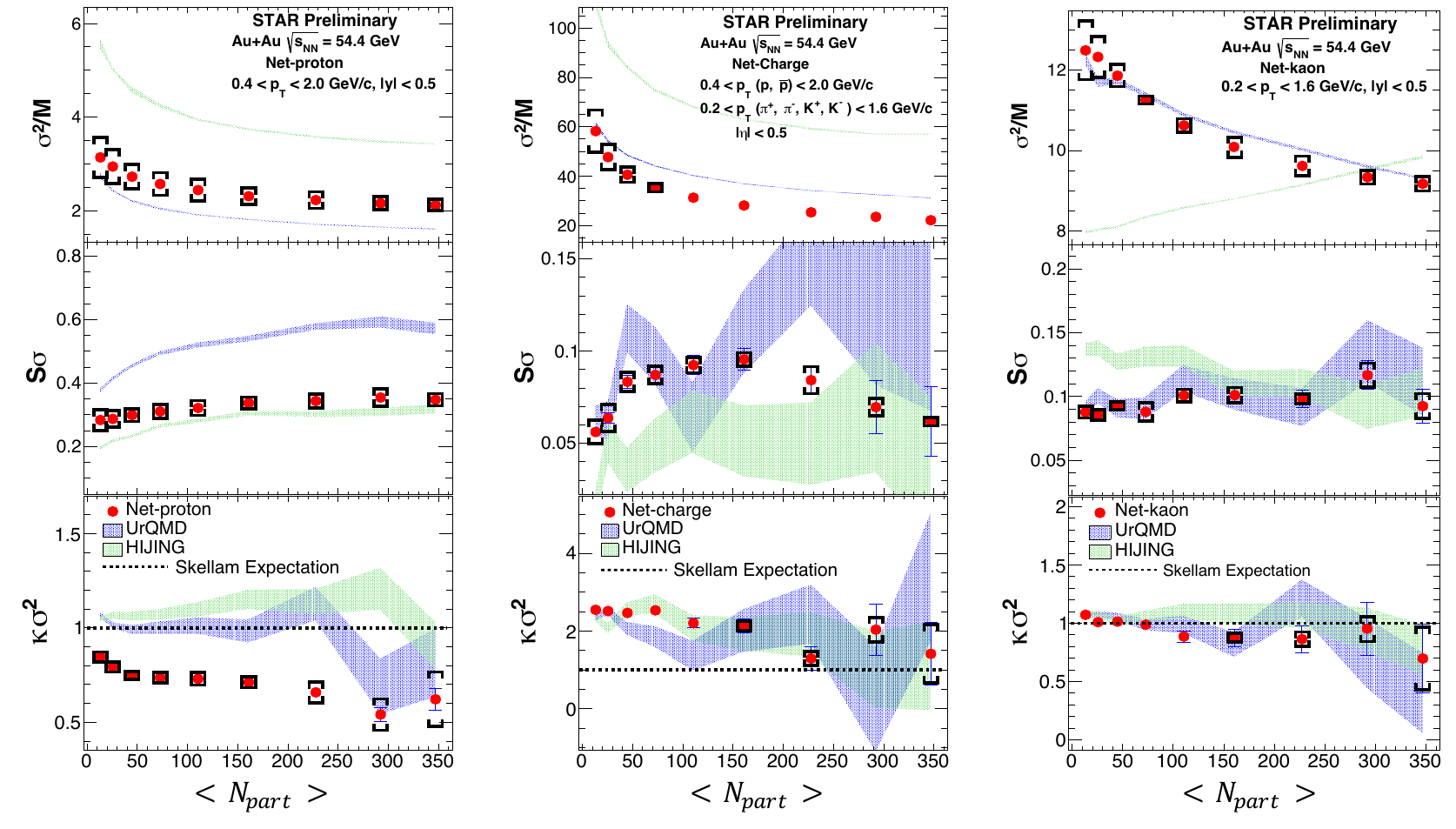}
 	\caption{Cumulant ratios $C_{2}/C_{1}$, $C_{3}/C_{2}$ and $C_{4}/C_{2}$ of net-proton, net-charge and net-kaon distributions as a function of average number of participant nucleons for Au+Au collisions at $\sqrt{s_{NN}}$ = 54.4 GeV. The blue and green band presents the UrQMD and HIJING model expectations caculated in the STAR acceptance, respectively.}
	 	\label{dedww}
\end{figure}
\subsection{Beam energy dependence of $C_{4}/C_{2}$ of net-particle distributions}
Beam-energy dependence of the cumulant ratio $C_{4}/C_{2}$ of net-particle distributions for peripheral (70-80\%) and most central (0-5\%) collisions with inclusion of the results from the current measurement (open and solid red markers respectively) are shown in the Fig. 3. Non-monotonic beam energy dependence of $C_{4}/C_{2}$ is observed for net-proton distribution while  $C_{4}/C_{2}$ of net-charge and net-kaon distributions show monotonic variation as a function of beam energy. The new $C_{4}/C_{2}$ measurements of net-particle distributions at $\sqrt{s_{NN}}$ = 54.4 GeV are consistent with the trends established for the other BES energies.
 \begin{figure}[h]
 	\hspace*{+1.9cm}
	\includegraphics[scale=0.43]{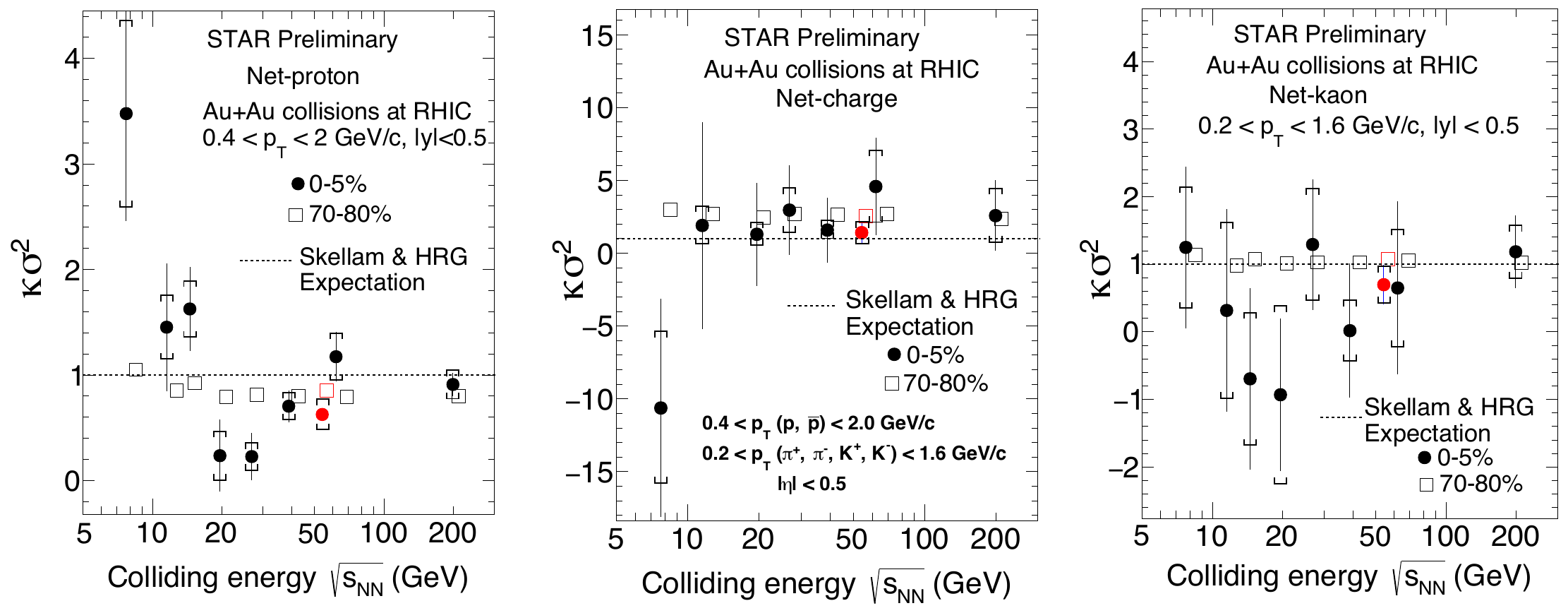}
	 \caption{Beam-energy dependence of $C_{4}/C_{2}$ for net-proton, net-charge and net-kaon distributions for 0-5\% and 70-80\%  most central collisions for Au+Au collisions with inclusion of the results from the  $C_{4}/C_{2}$ measurement at $\sqrt{s_{NN}}$ = 54.4 GeV (red markers).}
\end{figure}
\subsection{The sixth-order cumulant }
The collision-centrality dependence of the ratio of the sixth- to second-order cumulants ($C_{6}/C_{2}$)  of net-proton and net-charge distribution for Au+Au collisions are shown in Fig 4. The $C_{6}/C_{2}$ of net-proton distributions for central collisions (0-40\%) at $\sqrt{s_{NN}}$ = 54.4 GeV is positive while $C_{6}/C_{2}$ measured at $\sqrt{s_{NN}}$ = 200 GeV is negative for the same collision centrality~\cite{tnonaka}. A negative  $C_{6}/C_{2}$ is predicted for the crossover phase transition between hadronic matter and quark-gluon plasma in QCD-based calculations~\cite{negc6:friman}. The UrQMD model expectations for Au+Au collisions at $\sqrt{s_{NN}}$ = 200 GeV and $\sqrt{s_{NN}}$ = 54.4 GeV are found to be positive and consistent with the Skellam baseline across all collision centralities. The $C_{6}/C_{2}$ of net-charge distribution for Au+Au collisions at $\sqrt{s_{NN}}$ = 54.4 GeV for two most central collision classes (0-5\% and 5-10\% ) show positive sign. 
 \begin{figure}[h]
	\hspace*{+2.2cm}
	\includegraphics[scale=0.48]{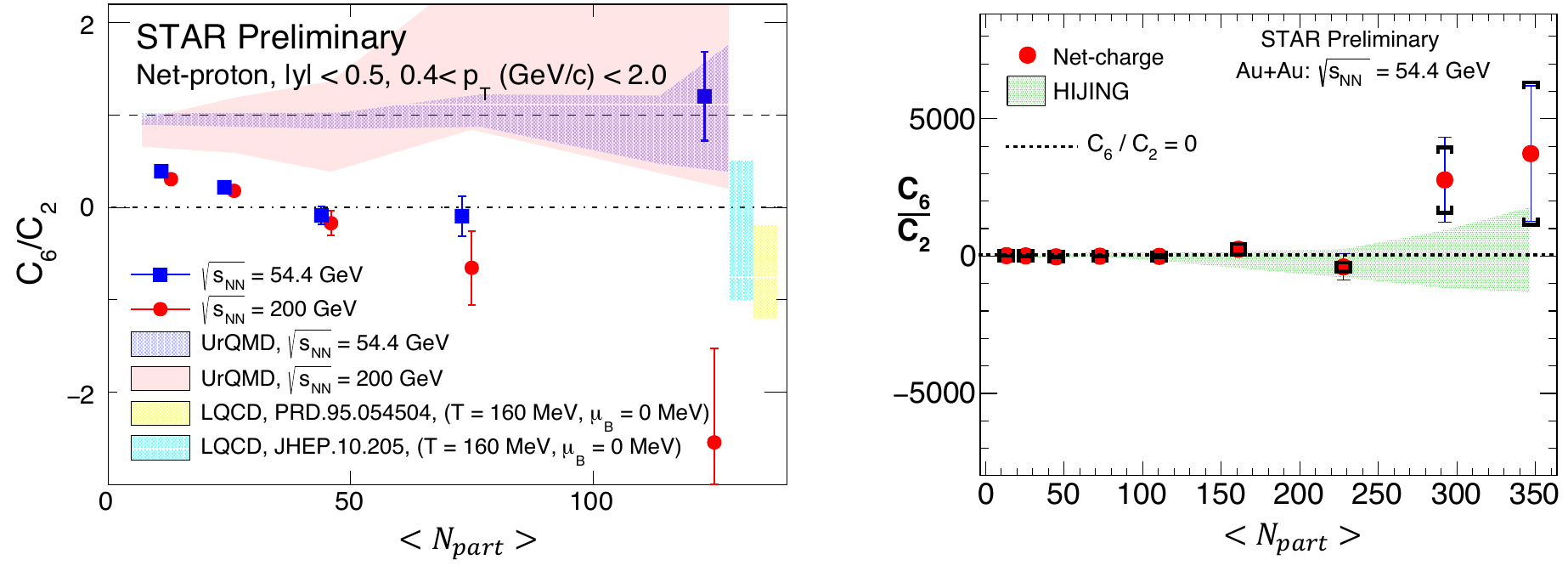}
	\caption{(Left plot) The $C_{6}/C_{2}$ of net-proton distribution for Au+Au collisions at $\sqrt{s_{NN}}$ = 54.4 GeV (blue) and 200 GeV (red) as a function of collision centrality. Red and blue bands are the UrQMD expectations for Au+Au collisions at $\sqrt{s_{NN}}$ = 200 and 54.4 GeV, respectively. The yellow and turquoise band are the Lattice QCD predictions. (Right plot) The $C_{6}/C_{2}$ of net-charge distribution for Au+Au collisions at $\sqrt{s_{NN}}$ = 54.4 GeV as a function of collision centrality. Green band is the HIJING model expectation.}
\end{figure}

\section{Summary}
We presented the collision-centrality dependence of cumulants and cumulant ratios of net-particle distributions from high statistics Au+Au collisions at $\sqrt{s_{NN}}$ = 54.4 GeV. The cumulants ($C_{n}, n$ $\leq$ 4) of net-particle distributions increase with average number of participant nucleons whereas the cumulant ratios $C_{3}/C_{2}$, $C_{4}/C_{2}$ and $C_{6}/C_{2}$ exhibit a weak collision-centrality dependence. The $C_{4}/C_{2}$ ($\kappa \sigma^{2}$) measurement of net-particle distributions at $\sqrt{s_{NN}}$ = 54.4 GeV are consistent with the existing beam energy trend of $C_{4}/C_{2}$ from the BES-I program. Furthurmore, the $C_{6}/C_{2}$ of net-proton distribution for Au+Au collisions at $\sqrt{s_{NN}}$ = 200 GeV for most central collisions (0-40\%) is negative, and this could be an experimental evidence of crossover phase transition. The new measurements at $\sqrt{s_{NN}}$ = 54.4 GeV serve precise baselines for the critical fluctuation studies which are expected at lower center-of-mass energies. 
\section{Acknowledgments}
We acknowledge the financial support by Department of Atomic Energy, Govt. of India.











\begin{thebibliography}{6}

\bibitem{Aoki:2006we} 
Y.~Aoki, G.~Endrodi, Z.~Fodor, S.~D.~Katz and K.~K.~Szabo, Nature {\bf 443}, 675 (2006).
\bibitem{Ejiri:2008xt} 
S.~Ejiri, Phys.\ Rev.\  {\bf D78}, 074507 (2008).
\bibitem {Stephanov:2008qz}
M.~A.~Stephanov, Phys.\ Rev.\ Lett.\  {\bf 102}, 032301 (2009).
\bibitem {Gavai:2010zn}
R.~V.~Gavai and S.~Gupta, Phys.\ Lett.\ B {\bf 696}, 459 (2011).
\bibitem {Gupta:2011wh}
S.~Gupta, X.~Luo, B.~Mohanty, H.~G.~Ritter and N.~Xu, Science {\bf 332}, 1525 (2011).
\bibitem{starNC} L.~Adamczyk {\it et al.} [STAR Collaboration], Phys.\ Rev.\ Lett.\  {\bf 113}, 092301 (2014). 
\bibitem{starNK} L.~Adamczyk {\it et al.} [STAR Collaboration], Phys.\ Lett.\  {\bf B785}, 551 (2018).
\bibitem{starNP_old} M.~M.~Aggarwal {\it et al.} [STAR Collaboration], Phys.\ Rev.\ Lett.\  {\bf 105}, 022302 (2010).
\bibitem{starNP} L.~Adamczyk {\it et al.} [STAR Collaboration], Phys.\ Rev.\ Lett.\  {\bf 112}, 032302 (2014).
\bibitem{shor_nature} J. Adam {\it et al.}, [STAR Collaboration], arXiv:2001.02852.
\bibitem {negc6:friman} B. Friman {\it et al.}, Eur. 346 Phys. J. {\bf C71}, 1694 (2011).
\bibitem{Shao:2005iu} 
M.~Shao {\it et al.}, Nucl.\ Instrum.\ Meth.\  {\bf A558}, 419 (2006).
\bibitem{arghya_auto} A. Chatterjee, Y. Zhang, J. Zeng, N. R. Sahoo, X. Luo, arXiv:1910.08004.
\bibitem{bin_width} X.~Luo, J.~Xu, B.~Mohanty and N.~Xu, J.\ Phys.\  {\bf G40}, 105104 (2013).
\bibitem{efcor_luo} X.~Luo, Phys. Rev.  {\bf C91}, 034907 (2015).
\bibitem{Nonaka:2017kko} T.~Nonaka, M.~Kitazawa and S.~Esumi, Phys.\ Rev.\ {\bf C95}, no. 6, 064912 (2017).
\bibitem{Luo:2011tp} X.~Luo, J.\ Phys.\  {\bf G39}, 025008 (2012).
\bibitem{bootstrap} A.~Pandav, D.~Mallick and B.~Mohanty, Nucl. Phys. {\bf A991} (2019) 121608.
\bibitem{urqmd}S. A. Bass {\it et al.}, Prog. Part. Nucl. Phys. {\bf 41}, 255 (1998).
\bibitem{hijing} M.~Gyulassy {\it et al.}, Comput.\ Phys.\ Commun.\  {\bf 83}, 307 (1994).
\bibitem{tnonaka} T. Nonaka, [STAR collaboration], contribution to these proceedings. 

\end{thebibliography}
\end{document}